\documentstyle[mncite,psfig]{mn}

\def\et{\em et al. \em}
\def\msun{M_{\odot}}
\def\gte{\,\lower.6ex\hbox{$\buildrel >\over \sim$} \, }
\def\lte{\,\lower.6ex\hbox{$\buildrel <\over \sim$} \, }

\title{The ellipticities of Galactic and LMC globular clusters}

\author[S.P.Goodwin]{Simon P. Goodwin\\
            Astronomy Centre, University of Sussex, 
            Falmer, Brighton BN1 9QH, UK.}

\date{}

\begin{document}

\maketitle

\begin{abstract}

The correlations between the ellipticity and the age and mass of LMC globular 
clusters are examined and both are found to be weak.  It is concluded that 
neither of these properties are mainly responsible for the observed 
differences in the LMC and Galactic globular cluster ellipticity 
distributions.  Most importantly, age cannot be the primary factor 
in the LMC-Galaxy ellipticity differences, even is there is a relationship, 
as even the oldest LMC clusters are more elliptical than their Galactic 
counter-parts.

The strength of the tidal field of the parent galaxy is proposed as the 
dominant factor in determining the ellipticities of that galaxy's globular 
clusters.  A strong tidal field rapidly destroys velocity anisotropies in 
initially triaxial, rapidly rotating elliptical globular clusters.  A weak 
tidal field, however, is unable to remove these anisotropies and the clusters 
remain close to their initial shapes.

\end{abstract}

\begin{keywords}

globular clusters: general
 
\end{keywords}

\section{Introduction}

It has been observed that the globular clusters of the LMC (Large Magellanic 
Cloud) are significantly more elliptical than their Galactic 
counter-parts (Geisler \& Hodge 1980; Frenk \& Fall 1982; van den Bergh 1983; 
van den Bergh \& Morbey 1984; Kontizas \et 1989; Han \& Ryden 1994).  
Indeed, there does appear to be a general and significant difference 
between globular cluster ellipticities according to the morphology of the 
parent galaxy (Han \& Ryden 1994).

Kolmogorov-Smirnov tests of the ellipticity differences of the Galactic 
and LMC populations show that, at the 99.2\% confidence level, the globular 
clusters have been drawn from different parent populations  (the two 
distribtions are illustrated in Fig. 1).  In addition,  
there appears to be strong evidence that the Galactic globular clusters are 
oblate spheroids compared to an apparently triaxial LMC population (Han 
\& Ryden 1994).

Studies of the ellipticities of the LMC globular clusters indicate that 
they correlate with luminosity/mass (van den Bergh 1983; van den Bergh \& 
Morbey 1984; Kontizas \et 1989).  There is some debate in the literature 
upon the existence of a correlation of ellipticity with age within the 
LMC.  Frenk \& Fall (1982) discovered an apparent correlation, however 
van den Bergh \& Morbey (1984) argued that this relationship would be effected 
by foreground absorption and no significant relationship exists.  Kontizas 
\et (1989) also found only a weak correlation with age. 

This paper examines the correlations present in the LMC globular cluster 
system.  The main question that is addressed is the origin of the differences 
between the LMC and Galactic populations: are these differences in 
ellipticity due to dynamical evolution and of what sort?

\section{Analysis}

A sample of globular clusters is analysed in an attempt to 
investigate the differences between LMC and Galactic populations.  A 
sample of 25 LMC globular clusters has been collected from the literature, 
this is a smaller sample than used in a number of other studies due to the 
lack of clusters for which a number of different parameters have been well 
determined.  

\begin{figure}
\centerline{\psfig{figure=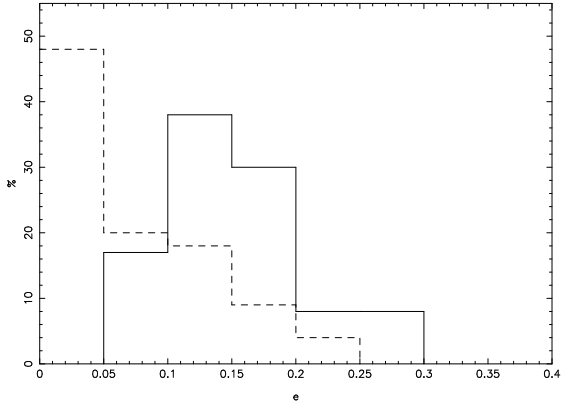,height=6.0cm,width=8.0cm,angle=270}}
\caption{The ellipticity distributions of globular clusters in the LMC 
(full line) and Galactic (dashed line) from data in White \& Shawl (1987) 
and Kontizas \et (1989).}
\label{figure:edist}
\end{figure}

Table 1 summarises the attributes of the clusters: $e_{\rm h}$ is the 
ellipticity measured at the half-mass radius, $M$ the mass, and $S$ the age 
parameter defined by Elson \& Fall (1985).  It should be noted that the $S$ 
parameter has a one-to-one relationship with the age type determinator used 
by Frenk \& Fall (1982) and van den Bergh \& Morbey (1984).

\begin{table}
\begin{center}
\begin{tabular}{|c|c|c|c|} \hline
NGC & $e_{\rm h}$ $^{\rm a}$ & $M/10^5 \msun$ & $S$ $^{\rm a}$ \\ \hline
1711 & 0.22 & 2.2 $^{\rm e}$ & 20 \\
1751 & 0.15 & 0.9 $^{\rm e}$ & 42 \\
1755 & 0.13 & 1.2 $^{\rm b}$ & 24 \\
1786 & 0.12 & 1.6 $^{\rm e}$ & 48 \\
1806 & 0.12 & 0.9 $^{\rm e}$ & 40 \\
1835 & 0.17 & 1.5 $^{\rm e}$ & 47 \\
1846 & 0.23 & 1.3 $^{\rm b}$ & 40 \\
1847 & 0.20 & 1.5 $^{\rm e}$ & 21 \\
1850 & 0.09 & 3.4 $^{\rm e}$ & 21 \\
1852 & 0.09 &                & 45 \\
1854 & 0.12 & 1.8 $^{\rm e}$ & 25 \\
1856 & 0.16 & 2.0 $^{\rm e}$ & 30 \\
1861 & 0.14 & 1.9 $^{\rm e}$ &    \\
1885 & 0.13 & 0.6 $^{\rm b}$ & 28 \\
1898 & 0.10 & 0.6 $^{\rm b}$ & 50 \\
1903 & 0.10 & 1.0 $^{\rm b}$ & 23 \\
1917 & 0.10 & 1.2 $^{\rm b}$ & 39 \\
1953 & 0.16 & 1.5 $^{\rm e}$ & 29 \\
1987 & 0.16 &                & 35 \\
2004 & 0.20 & 0.4 $^{\rm b}$ & 15 \\
2019 & 0.20 & 1.6 $^{\rm e}$ & 46 \\
2031 & 0.21 & 0.3 $^{\rm b}$ & 27 \\
2038 & 0.16 & 1.3 $^{\rm e}$ &    \\
2056 & 0.13 & 1.5 $^{\rm e}$ & 31 \\
2107 & 0.12 & 0.9 $^{\rm b}$ & 32 \\ \hline
\end{tabular}
\caption{Comprehensive data for 25 LMC globular clusters available in the 
literature.  $^{\rm a}$ from Kontizas \et (1989), $^{\rm b}$ from 
Chrysovergis, Kontizas \& Kontizas (1989), $^{\rm c}$ from Elson (1992), 
$^{\rm d}$ from Mateo (1987) and $^{\rm e}$ from Kontizas, Chrysovergis 
\& Kontizas (1987).}
\end{center}
\label{table:data}
\end{table} 

\begin{figure}
\centerline{\psfig{figure=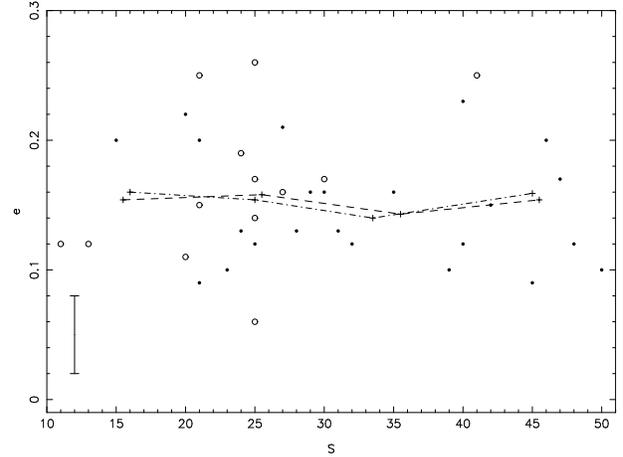,height=6.0cm,width=8.0cm,angle=270}}
\caption{Ellipticity-age relationship for LMC globular clusters.  Filled 
circles are data from table~\ref{table:data} while open circles are other 
clusters for which $e_{\rm h}$ and $S$ are given in Kontizas \et (1989). 
The estimated errors of $\pm 0.03$ are illustrated by the error 
bar in the lower left corner.  The 
dotted line shows the mean ellipticities in 4 age bins, and the dash-dot 
line shows the mean ellipticities in 4 numerical bins.}
\label{figure:e-a}
\end{figure}

Figure~\ref{figure:e-a} shows the age ($S$ parameter) against ellipticity for 
23 of the globular clusters in the sample.  The lack of significant 
correlation is obvious even by eye.  The lines on fig.~\ref{figure:e-a} show 
the mean ellipticities when the data is placed in 4 bins by age (dotted line) 
or by number of clusters (dash-dot line).  A statistical test using the sample 
correlation coefficient shows only a very weak correlation. 

The strong correlation found by Frenk \& Fall (1982) is highly dependent upon 
foreground absorption increasing the ellipticities of a number of their 
younger clusters.  Once this effect has been removed the correlation 
disappears (van den Bergh \& Morbey 1984).  Such a correlation is also 
absent in the more uniform sample of Kontizas \et (1989). 

It should be noted that ellipticity is not a simple quantity.  Ellipticity 
varies with radius (Kontizas \et 1989; Kontizas \et 1990) and so is 
difficult to define.  The half-mass radius ellipticities of Kontizas \et 
(1989) have been used as they form a more consistent set.  The use of 
ellipticities from Frenk \& Fall (1982), however, would produce an  
ellipticity-age relationship (although van den Bergh \& Morbey 1984 argue 
that it is not significant).  The existence of a stronger relationship should   
not be discounted entirely, although it would not appear to be the 
dominant relationship.

\begin{figure}
\centerline{\psfig{figure=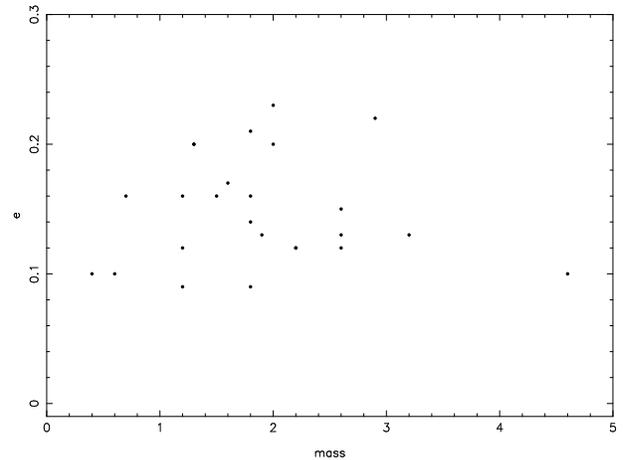,height=6.0cm,width=8.0cm,angle=270}}
\caption{Ellipticity-mass relationship for LMC globular clusters.}
\label{figure:e-m}
\end{figure}

The ellipticity-mass relationship illustrated in fig.~\ref{figure:e-m} also 
shows a surprising lack of correlation in contradiction to the results of van 
den Bergh \& Morbey (1984) and Kontizas \et (1989).  This figure is 
similar to fig. 5 of Kontizas \et (1989) but the inclusion of masses for 
NGCs 2004 and 2031 (from Chrysovergis \et 1989) help remove a correlation.  
Again the sample correlation coefficient test upon the data shows only a 
weak correlation.

There also appears to be no significant correlation between the age and 
mass of LMC globular clusters.  It would appear as if neither age nor mass are 
the dominant factors in the differences in ellipticities between the LMC 
and Galactic globular clusters.  

\section{Discussion} 

The main differences between the LMC and Galactic globular clusters may 
be summarised briefly as: 

\begin{enumerate}

\item The LMC globular clusters {\em at all ages} are significantly more 
elliptical than the Galactic population.

\item The shapes of the LMC (and SMC) globular clusters are well-represented by 
triaxial spheroids while those of the Galaxy (and M31) are oblate spheroids 
(Han \& Ryden 1994)

\end{enumerate}

Han \& Ryden (1994) attribute (ii) to age differences between the two 
populations, the older Galactic population being more relaxed and hence 
more spherical then the younger LMC population.  However, this cannot be 
the entire picture as (i) indicates that even old LMC clusters are 
more elliptical and triaxial than coeval Galactic globular clusters. 

In most other respects the Galactic and LMC globular cluster populations 
are very similar.  The LMC globular clusters have masses in the range 
$10^4$ to a few $\times 10^5 \msun$ (Elson, Fall \& Freeman 1987; 
Chyrsovergis \et 1989; Lupton \et 1989) and core radii of $0.5 < r_{\rm c}/{\rm pc} < 6.8$ (Elson 1991).  These values are very similar to those of the 
Galactic globular clusters $10^4 < M/\msun < 10^6$ (Mandushev, Staneva 
\& Spansova 1991) and $0.1 < r_{\rm c}/{\rm pc} < 19$ (Djorgovski 
1993).

Young globular clusters appear always to show high ellipticity. In addition 
to the LMC, recent HST observations of super star clusters (assumed analogous 
to young globular clusters) in NGC 1569 also appear to show high ellipticities 
(O'Connell, Gallagher \& Hunter 1994).  If high ellipticity, and 
presumably a triaxial shape, is a general property of a young globular 
cluster, a question must be raised as to why the old Galactic population 
have had their original structures modified while the old LMC population 
remain unchanged, especially as they appear so similar in most other 
respects.

This being the case some dynamical influence must be invoked to explain the 
differences in the populations, it is proposed that this factor is the 
strength of the tidal field of the parent galaxy.  If globular clusters 
form as triaxial spheroids (with a suitably anisotropic velocity 
dispersion) then the action of a strong tidal field as they orbit about a 
galaxy will be to force the velocity dispersion to isotropy.  In other 
words, the cluster will loose its initial triaxiality and become 
more spherical.

The simulations of Longaretti \& Lagoute (1996a,b) of rotating globular 
clusters also show the effects of a strong tidal field in reducing the 
half-mass ellipticities of globular clusters, by removing angular momentum 
from the cluster.  Globular clusters experiencing a higher tidal force 
(lower Galactocentric radius) become spherical more rapidly than other 
clusters (figs. 7e and 8e in Longaretti \& Lagoute 1996a).

If the tidal field is dominant in reducing ellipticity then the 
observed dependence of ellipticity upon galaxy morphology (Han \& Ryden 1994) 
would be expected.  Further, the slightly higher ellipticities observed in 
the SMC than in even the LMC (Kontizas \et 1990, Han \& Ryden 1994) would 
also be expected due to the weaker tidal field of the SMC.

One might still expect an ellipticity-age relationship in this scenario (of 
the sort noted by Frenk \& Fall 1982).  The apparent absence of such a 
relationship is presumably due to the inability of the LMC tidal field 
to modify the shapes of its globular cluster population significantly.  
Galactic globular clusters are presumably all old enough to have had their 
initial triaxiality destroyed and their shapes are mainly due to rotation 
(White \& Shawl 1987) and, possibly, shocking.  
The most elliptical Galactic globular clusters are found to have low 
Galactocentric radii and (for all but 2 clusters) low heights above the 
Galactic disc (using data from White \& Shawl 1987 and Djorgovski 
1993).  This may be explained as the result of recent tidal shocking (disc or 
bulge) reintroducing an anisotropy into the velocity dispersion.  This may 
account for the large spread of ellipticities at low Galactocentric radii 
($0.00 < e < 0.27$ at $R_{\rm G} < 6$kpc). 

\section{Summary}

The tidal field of the parent galaxy is proposed as the major factor 
determining the ellipticities of its globular cluster population.  If the 
original shapes of all globular cluster populations are highly elliptical and 
triaxial then a strong tidal field will act to reduce that triaxiality and 
force the clusters to a more isotropic distribution and spherical shape.  The 
effectiveness of this process is limited in weak tidal fields (such as those 
of the LMC and SMC) leading to the ellipticities of even the oldest 
globular clusters remaining at their high initial values. 

\section{Aknowledgements}

S.\,P.\,Goodwin is a DPhil student at the University of Sussex currently in 
reciept of a PPARC grant.

\end{document}